\documentclass[a4paper]{jpconf}
\usepackage{graphicx}
\usepackage{array}
\usepackage{amssymb}

\usepackage{color}
\definecolor{cyan}{rgb}{0.2,0.6,1}
\definecolor{red}{rgb}{1,0,0}
\definecolor{blue}{rgb}{0,0,1}
\definecolor{green}{rgb}{0,0.5,0}

\begin{document}
\title{Influence of the Anderson Transition on Thermoelectric Energy Conversion in Disordered Electronic Systems}

\author{Ilia Khomchenko$^{1, 2, 3}$, Henni Ouerdane$^1$, Giuliano Benenti$^{1, 2, 4}$}

\address{$^1$Center for Digital Engineering, Skolkovo Institute of Science and Technology, 30 Bolshoy Boulevard, bld. 1, 121205, Moscow, Russia}

\address{$^2$ Center for Nonlinear and Complex Systems, Dipartimento di Scienza e Alta Tecnologia, Università degli Studi dell’Insubria, via Valleggio 11, 22100 Como, Italy}

\address{$^3$ Istituto Nazionale di Fisica Nucleare, Sezione di Milano, via Celoria 16, 20133 Milano, Italy}

\address{$^4$ NEST, Istituto Nanoscienze-CNR, I-56126 Pisa, Italy}

\ead{ikhomchenko@uninsubria.it, H.Ouerdane@skoltech.ru, giuliano.benenti@uninsubria.it}

\begin{abstract}
So far, the efficiency of thermoelectric energy conversion remains low compared to traditional technologies, such as coal or nuclear. This low efficiency can be explained by connecting the thermoelastic properties of the electronic working fluid to its transport properties. Such connection also shows that operating close to electronic phase transitions can be an efficient way to boost the thermoelectric energy conversion. In this paper, we analyze thermoelectric efficiency close to the metal-insulator Anderson transition. Our results reveal the direct link between the thermoelectric and thermoelastic properties of Anderson-type systems. Moreover, the role of the conductivity critical exponent in the thermoelectric energy conversion is analysed. Finally, we show that relatively large values of the thermoelectric figure of merit may be obtained in the vicinity of the Anderson transition. 
\end{abstract}

\section{Introduction}
Thermoelectric conversion performance is usually determined using a combination of three transport coefficients: the electrical conductivity $\sigma$, the thermal conductivity $\kappa$, and the Seebeck coefficient $\alpha$. This combination is known as the thermoelectric figure of merit $zT$~\cite{thermoelements1957af,goldsmid2016introduction,benenti17}:

\begin{equation}
zT = \frac{\alpha^{2}\sigma}{\kappa}T,
\label{eq:1.1}
\end{equation}
where $T$ is the average temperature across the system. As phonons of the crystal lattice and charge carriers $-$ usually electrons, contribute to thermal transport, $\kappa$ can be written as $\kappa = \kappa_{\rm e} + \kappa_{\rm ph}$. 

The range of applications of thermoelectric technology would be significantly extended, provided $zT$ exceeds a value of at least $4$~\cite{vining2009inconvenient} and that devices operate under appropriate working conditions \cite{Apertet2012}. This is a formidable challenge, due to the interdependence of the transport coefficients, ruled by phenomenological laws such as the Wiedemann–Franz law connecting heat and electric conductivity~\cite{ashcroft1976solid}. To establish an upper bound for $zT$ under given working conditions, it is enough to consider the thermoelectric figure of merit of the conduction electron gas alone, $z_{\rm e}T$, which disregards the lattice thermal conductivity $\kappa_{\rm ph}$:    
\begin{equation}
z_{\rm e}T = \frac{\alpha^{2}\sigma}{\kappa_{\rm e}}T > zT
\label{eq:1.2}
\end{equation}

Coupling between heat and electrical transport, as was underlined by Apertet et al.~\cite{apertet2012internal}, results in a convective process, namely a heat flow associated with the displacement of charge carriers. The convective part of the heat flow, which adds to the conductive part due to electrons and phonons under open circuit conditions, can be enhanced near the critical temperature of an electronic transition~\cite{vining1997thermoelectric, ouerdane2015enhanced}. In the literature, thermoelectric conversion was discussed for superconducting~\cite{khomchenko2022} and metal-insulator Anderson transition~\cite{imry2010localization,benenti2016thermoelectric, yamamoto2017thermoelectricity,Garmroudi2022}.

To address the conversion efficiency optimization problem, it is instructive to analyze the thermodynamic properties of the conduction electron gas, which is the actual working fluid of thermoelectric devices. The Seebeck coefficient given by the ratio of the gradients of two intensive variables, electrochemical potential $\mu$ and temperature $T$: $\alpha = -\nabla \mu/(q\nabla T)$ \cite{Apertet2016}, has a thermostatic counterpart: $\alpha_{\rm th} = -{\rm d}\mu/{\rm d}T$, which derives from the Gibbs-Duhem equation, as $\alpha_{\rm th} = S/N$ is the entropy per particle. The quantity $\alpha_{\rm th}$ can be written using also the definitions of the thermoelastic coefficients of the electron gas and the Maxwell relations, and a thermodynamic figure of can be defined from the calculation of the isentropic expansion factor~\cite{ouerdane2015enhanced}:
\begin{equation}
Z_{\rm th}T= \frac{\beta^{2}}{\chi_{T}C_{N}}T,
\label{eq:1.3}
\end{equation}
where $\beta$ is the analogue to thermal dilatation coefficient, $\chi_{T}$ is the analogue to isothermal compressibility, and $C_{N}$ is the analogue to specific heat at constant volume. Definitions are given further below.

As discussed in \cite{ouerdane2015enhanced,khomchenko2022}, driving the electron gas close to a phase transition yields a significant increase of the isentropic expansion factor, which boosts the energy conversion efficiency. In the latter works, thermally driven effects, namely fluctuating Cooper pairs and nematic fluctuations, were considered in 2D systems and thin films. Other effects such as disorder, can influence the thermodynamic and transport properties of the electron gas: in a disordered system, the charge carriers states at a given energy can either be localized or delocalized depending on the disorder strength. In this work, we analyze the effects of the transition from the delocalized to localized states, or the Anderson metal-insulator transition, on the thermodynamic properties of the electron gas and its ability to perform an efficient energy conversion in the vicinity of the critical point. Indeed, one may expect that the Seebeck coefficient may drastically increase as the system is driven away from its metallic phase as the entropy per carrier increases.
Thermoelectric conversion near the Anderson transition has been studied in~\cite{imry2010localization, benenti2016thermoelectric, yamamoto2017thermoelectricity}, but the link between thermoelastic and transport properties has not been yet considered, while it has been studied for the metal to superconductor phase transition~\cite{khomchenko2022}.

The paper is organized as follows. In the next two sections, for completeness and clarity, we give a brief recap of the basic ingredients of our approach: the transport coefficients and the thermoelastic coefficients. We then focus on the Anderson transition, detailing the assumptions and parameters we use for our model. We present and discuss our numerical results in the subsequent section, and we end the paper with concluding remarks.

\section{Transport coefficients}
The standard approach to calculate $\sigma$, $\alpha$, and $\kappa_{\rm e}$ is to relate these transport coefficients to Onsager's kinetic coefficients $L_{ij}$ $(i,j=1,2)$, in the frame of linear non-equilibrium thermodynamics~\cite{onsager1931reciprocal1,goupil2011thermodynamics, pottier2012physique}: 

\begin{eqnarray}
\label{eq:2.1}
 &&  \sigma = \frac{e^{2} L_{11}}{T} , \\
 \label{eq:2.2}
 && \alpha =  \frac{L_{12}}{eT L_{11}},  \\
 \label{eq:2.3}
 && \kappa_{\rm e} = \frac{1}{T^{2}}\left[{L}_{22} - \frac{{L}_{12}{L}_{21}}{{L}_{11}} \right ], 
\end{eqnarray}
where $e$ is the electron charge. 

To compute the Onsager coefficients $L_{ij}$, we use the Boltzmann equation in the relaxation time approximation~\cite{mahan1996best}:

\begin{eqnarray}
\label{eq:2.4}
L_{11} & = & T\int_0^{\infty} \Sigma(E) \left(-\frac{\partial f}{\partial E}\right){\rm d}E,\\
\label{eq:2.5}
L_{12} = L_{21} & = & T\int_0^{\infty} (E - \mu)\Sigma(E) \left(-\frac{\partial f}{\partial E}\right){\rm d}E,\\
\label{eq:2.6}
L_{22} & = & T\int_0^{\infty} (E - \mu)^2\Sigma(E) \left(-\frac{\partial f}{\partial E}\right){\rm d}E,
\end{eqnarray}
where $f=\{\exp[(E-\mu)/k_{\rm B}T] + 1 \}^{-1}$ is the Fermi distribution function, $\Sigma(E)$ is the transport distribution function, and $\mu$ is the electrochemical potential introduced above. Here, $\Sigma(E) = \tau(E)v^2(E)g(E)$ is the transport distribution function, with $\tau(E)$ being the electron relaxation time, $v(E)$ the electron group velocity, and $g(E)$ the density of states. Note that the relaxation time depends on the model of the system and also varies with respect to the type of collisions~\cite{pottier2012physique}.

\section{Thermoelastic coefficients}
The thermodynamics of the noninteracting electron gas is very similar to its classical gas counterpart. Using the correspondence between volume $V$ and the number of electrons $N$ and pressure $P$ and the chemical potential $\mu$, namely, $V \longrightarrow N$ and $-P \longrightarrow \mu$, one can define analogous coefficients for the electron gas $\beta$, $\chi_T$, $C_{N}$ (already introduced in Eq.~(\ref{eq:1.3})), and $C_{\mu}$~\cite{ouerdane2015enhanced}, where $C_{\mu}$ is the analogue to specific heat at constant pressure.

Following the same approach as with transport coefficients, we relate thermoelastic coefficients to the distribution function~\cite{ouerdane2015enhanced, brantut2013thermoelectric}. The analogue to isothermal compressibility is:
\begin{equation}
\label{eq:3.1}
\chi_{T} N=\int_{0}^{\infty} g(E)\left(-\frac{\partial f}{\partial E}\right) {\rm d}E;
\end{equation}
the analogue of thermal dilatation coefficient reads: 
\begin{equation}
\label{eq:3.2}
\beta N=\frac{1}{T}\int_{0}^{\infty} g(E)(E-\mu)\left(-\frac{\partial f}{\partial E}\right) {\rm d}E;
\end{equation}
and the specific heat at constant electrochemical potential is given by:
\begin{equation}
\label{eq:3.3}
C_{\mu} =\frac{1}{T}\int_{0}^{\infty} g(E)(E-\mu)^{2}\left(-\frac{\partial f}{\partial E}\right) dE. 
\end{equation}
The specific heat at constant particle number $C_{N}$ and at constant electrochemical potential $C_{\mu}$ are connected via the Maxwell’s relation
\begin{equation}
C_{\mu} = C_N + \frac{\beta^2 T}{\chi_T} = C_N \left(1 + \frac{\beta^2}{\chi_T C_N}T\right).
\label{eq:3.4}
\end{equation}

\section{Anderson transition model}
Anderson developed the concept of localized and extended states with a simple theoretical model: if a quantum-mechanical system is sufficiently disordered (e.g., a semiconductor with lattice defects or impurities) and at sufficiently low carrier density, diffusion cannot take place, which results in wave function localization ~\cite{anderson1958absence}. The model assumes a distribution of sites occupied by particles, which may be random or might be regular in three-dimensional space. Later, Mott introduced the mobility edge concept \cite{Mott1967}, which is an energy level, $E_{\rm c}$, in a bulk semiconducting system's energy band, e.g., an impurity band, which separates localized states (with energy $E < E_{\rm c}$) from extended states (with energy $E > E_{\rm c}$) within the same band. The location of $E_{\rm c}$ within the band depends on the disorder strength (i.e. number and type of defects, impurities, etc.) and the density of states $g(E)$ near the mobility edge follows a power-law:
\begin{equation}\label{eq:4.2}
g(E) =  const \ |E-E_{\rm c}|^{(d-y)/y}, 
\end{equation}
\noindent where $d$ is the system's dimension, and $y$ a scaling parameter, and the expression for the electrical conductivity at zero temperature reads \cite{wegner1976electrons}:
\begin{equation}
\label{eq:4.3}
 \sigma(T=0,E) = \left\{ \begin{array}{lll} A (E-E_{\rm c})^{(d-2)/y}, & \mbox{if} & E \geq E_{c}, \\ 0, & \mbox{if} & E \leq E_{\rm c}, \end{array}\right.
\end{equation}
where $A$ is a constant. Moreover, the electrical conductivity at zero temperature is related to the transport distribution function as $\sigma(T=0, E)=2e^{2}\Sigma(E)$~\cite{benenti2016thermoelectric}. We are here only interested in the $d=3$ case because of the absence of quantum diffusion in two dimensions and one dimension~\cite{abrahams1979scaling, lancaster2014quantum} for which there is no mobility edge $E_{\rm c}$. The theory is valid for the values of $y$ in the range $0<y<d$~\cite{wegner1976electrons}.

In the present work, we consider the electron gas at the metal-insulator transition~\cite{lemarie2009observation}. The calculations below are done using Eq.~ (\ref{eq:4.2}), and with $g(E_{\rm F}) =2.45 \times 10^{24}$ J$^{-1}\cdot$m$^{-3}$ at the Fermi level. The constant $A$ in Eq.~(\ref{eq:4.3}) was chosen as $A=10^{22}$ $\Omega^{-1} \cdot$ m$^{-1} \cdot$ J$^{-x}$~\cite{enderby1994electron} to match the typical values of the electrical conductivity.  For convenience, let $x = 1/y$, which is also known as the conductivity critical exponent. We investigate the influence of this parameter on the thermodynamic and thermoelectric figure of merit, $Z_{\rm th}T$ and $z_{\rm e}T$, showing their dependencies for several values of $x$. We consider a range starting at $x=1/3$. Note that the exact value of $x$ is not known, notwithstanding the various numerical, analytical, and experimental methods used for determining it~\cite{imry2010localization}. For example, MacKinnon provides the value of $x=1.54$, which is in range of typical values $0.5<x<2$~\cite{mackinnon1994critical}. In our calculations, we use the following expressions for the temperature-dependent chemical potential of the three-dimensional electron gas reads~\cite{sotnikov2017chemical}:
\begin{equation}
\label{eq:4.4}
 \mu(T) = E_{\rm F}\left\{\begin{array}{lll} 1 + a_{1}t + a_{2}t^{2} + a_{3}t^{3}+a_{4}t^{4}, & \mbox{if} & 0 \leq t \leq t^{*}, \\  t \mathrm{ln}(b_{+} t^{-3/2})- t \mathrm{ln}(1 - b_{+} (2t)^{-3/2}), & \mbox{if} & t \geq t^{*}, \end{array}\right.
\end{equation}
where $t=T/T_{\rm F}$, with $T_{\rm F}=E_{\rm F}/k_{\rm B}$ being the Fermi temperature; $t^{*}=1.36$. The coefficient $b_{+}$ is given by $b_{+} = a_{+}/ \Gamma(3/2)$, $a_{+}=2/3$, $\Gamma(x)$ is the gamma function, $a_{1}=0.016$, $a_{2}=-0.957$, $a_{3}=-0.293$, and $a_{4}=0.209$~\cite{sotnikov2017chemical}. In our calculations, $T_{\rm F}=42.3$ K for the three-dimensional electron gas, whose concentration $n$ was taken as $n=10^{18} \mathrm{cm}^{-3}$. The mobility edge is set to $E_{\rm c}(T) = 0$ eV, based on the simplified model of Wegner~\cite{wegner1976electrons}.

\section{Results and discussion}
As we study the effect of disorder on thermoelectric energy conversion, several values of the critical exponent are considered. Our model includes extended and localized sates, with a smaller proportion of the last ones since the mobility edge $E_{\rm c} =0$. As for the extended states, their influence is significant because for the system with weak enough disorder, the extended states may exist and contribute to the conductivity, which is finite even at zero temperature~\cite{huckestein1995scaling}. 

\begin{figure}
\includegraphics[angle=0,origin=c, width=1\textwidth]{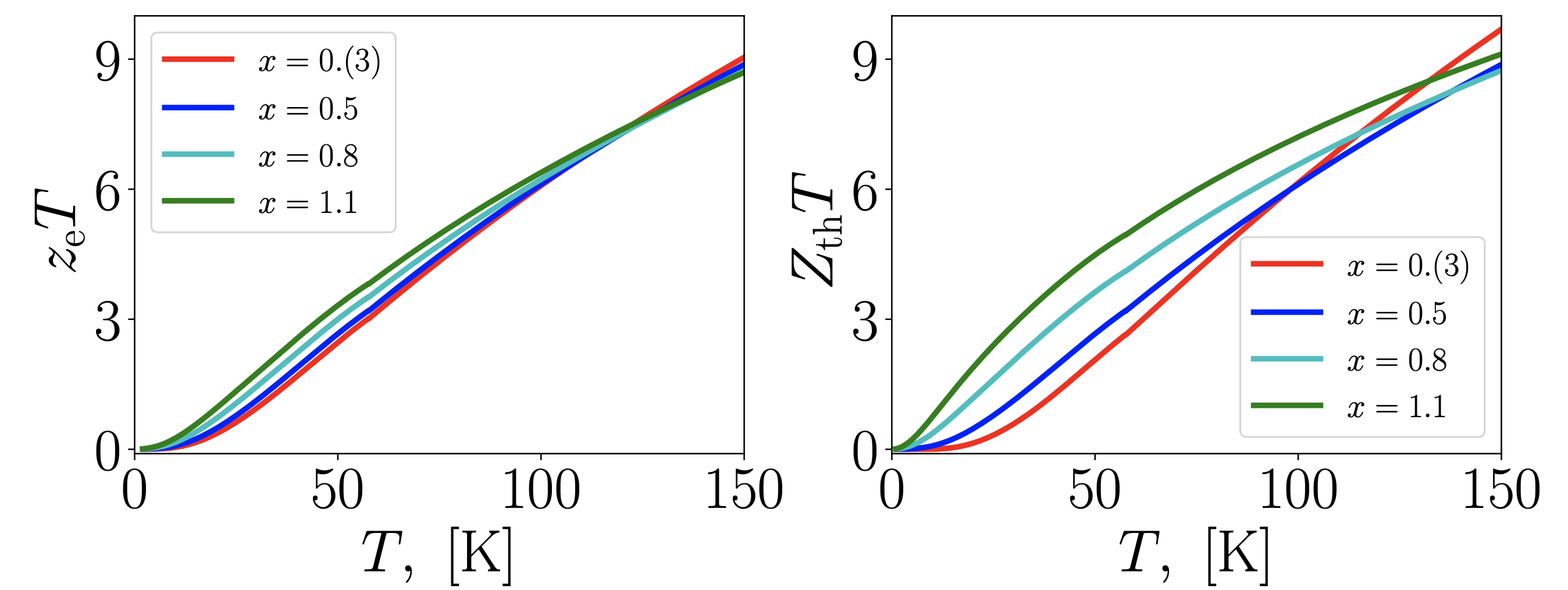}
\caption{\label{fig:1} The electron gas figure of merit $z_{\rm e}T$ (left) and the figure of merit $Z_{\rm th}T$ (right) are depicted functions of the temperature $T$, for different values of the critical exponent $x$. The notation $0.(3)$ refers to the repeating decimal $1/3=0.(3)$.}
\end{figure}

Figure~\ref{fig:1} illustrates the electron gas figure of merit $z_{\rm e}T$ and the thermodynamic figure of merit $Z_{\rm th}T$ as functions of the temperature $T$. To compute $z_{\rm e}T$, we calculated first the transport coefficients $\kappa_{e}$, $\sigma$, and $\alpha$ numerically integrating Eqs.~(\ref{eq:2.4}-\ref{eq:2.6}) using the transmission function $\Sigma(E) = \sigma(T=0,E)/2e^{2}$~\cite{benenti2016thermoelectric}, where $\sigma(T=0,E)$ is given by Eq.~(\ref{eq:4.3}). As regards the thermodynamic figure of merit $Z_{\rm th}T$, we integrated Eqs.~(\ref{eq:3.1},\ref{eq:3.2},\ref{eq:3.3}) using the density of states $g(E)$ given in Eq.~(\ref{eq:4.2}).

Both figures of merit grow steadily over the whole temperature range. Importantly, both $z_{\rm e}$ and $Z_{\rm th}$ increase as the system is driven close to the phase transition, implying that the increase of the electron gas isentropic expansion factors fosters the desired behavior of the transport parameters. At relatively small temperatures ($T \lesssim 150$ K), the larger the critical exponent, the larger the figures of merit, while at higher temperatures ($T \gtrsim 150$ K), the dependence is reversed. This shows that thermal excitation of carriers get them into the extended states, however the crossovers of the curves computed for various values of $x$ as $T$ increases shows that the interplay between disorder effects and temperature effects is not trivial. 

\begin{figure}
\begin{center}
\includegraphics[angle=0,origin=c, width=0.7\textwidth]{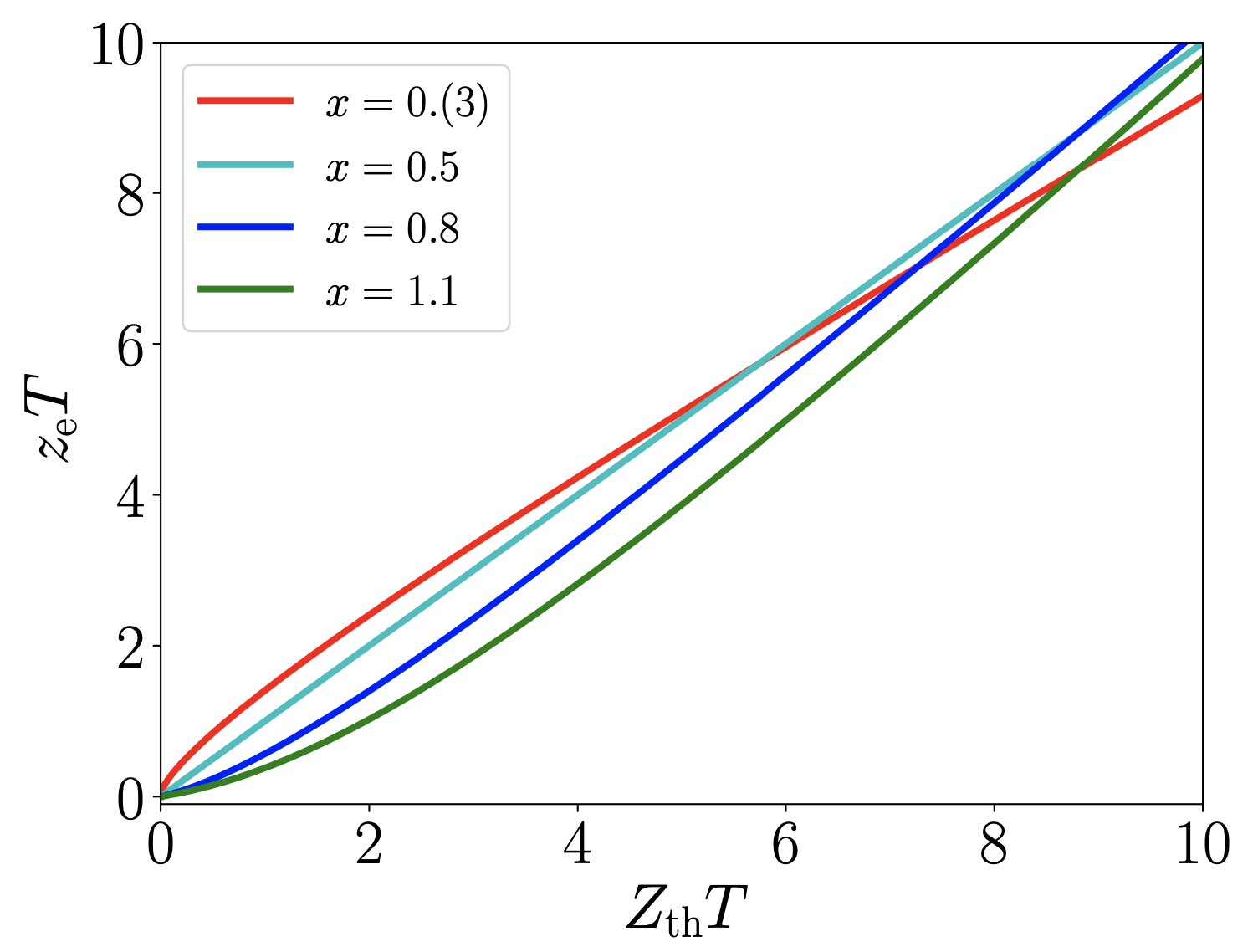}
\caption{\label{fig:2} The parametric plot of the figure of merit $z_{\rm e}T$ versus the figure of merit $Z_{\rm th}T$ for various values of the critical exponent.}
\vspace*{-0.4cm}
\end{center}
\end{figure}

To understand how thermoelastic properties are reflected in transport properties, we show the correlation between the figure of merit $z_{\rm e}T$ and the figure of merit $Z_{\rm th}T$ in Fig.~\ref{fig:2}. Depending on the value of the critical exponent, the correlation between the figures of merit varies from a straight line at $x=0.5$ to a power law behavior for other values of $x$. We underline that $z_{\rm e}T$ is a monotonously growing function of $Z_{\rm th}T$, regardless of the value of the critical exponent. This means that we can estimate how efficient energy conversion can be in the best-case scenario (that is, neglecting phonons) on purely thermodynamic grounds, by studying the behavior of the figure of merit $Z_{\rm th}T$ rather than $z_{\rm e}T$.
  
Next, we show how efficient is energy conversion near the transition temperature and in the whole temperature range. To characterize the efficiency of this process, we introduce the thermodynamic efficiency   
\begin{equation}
\eta_{\rm max} = \frac{\sqrt{\gamma}-1}{\sqrt{\gamma}+1} \eta_{\rm C}, 
\label{eq:5.1}
\end{equation}
\noindent where $\gamma = C_{\mu}/C_{N}$ is an analogue to the classical isentropic expansion factor, and $\eta_{\rm C}$ is the Carnot efficiency. At $T\approx 50$ K, when $k_B T\approx \mu-E_c$ and thermoelectric transport is affected by the non-analytical behavior of the transmission function $\Sigma(E)$ (see Eq.~(\ref{eq:4.3})), and $\eta_{\rm max}$ varies from around $0.28\eta_{\rm C}$ to roughly $0.4\eta_{\rm C}$ when the critical exponent $x$ varies from $0.(3)$ to $1.1$. For higher temperatures ($T \approx 150$ K), we can get the performance $\eta_{\rm max}\approx 0.5\eta_{\rm C}$. Of course, such high efficiency is only an upper bound, as phonons become more relevant with increasing temperature. Figures reported in Table~\ref{table:1} show how $\eta_{\rm max}/\eta_{\rm C}$ varies as a function of the critical exponent $x$. 

\begin{table}[ht!]
\caption{Dependence of the maximum thermodynamic efficiency on the critical exponent.} 
\begin{flushleft}
    \begin{center}
    \begin{tabular}{|p{1.5cm}|m{1cm}|m{1cm}|m{1cm}|m{1cm}|m{1cm}|m{1cm}|m{1cm}|m{1cm}|} 
    \hline
    {\bf $x$} & {0.(3)} & {0.5} & {0.8} & {1.1} & {1.4}& {1.7} & {2.0} & {2.3}\\
   \hline
    &~&~&~&~&~&~&~&~\\
    {\bf $\eta_{\rm max}/\eta_{\rm C}$}  & {0.28} & {0.32} & {0.37} & {0.4} & {0.43} & {0.46} & {0.48} & {0.49} \\
    \hline
\end{tabular}
\end{center}
\end{flushleft}
\label{table:1}
\end{table}

Though the trend shown in the table appears simple, its explanation is not, as establishing a clear relationship between the isentropic expansion factor and the critical exponent $x$ would require an analysis beyond the scope of the present work, involving a correlation between the thermoelastic coefficients of the electron gas in the extended state, the disorder strength and the critical exponent. Here, we may suggest the following interpretation. With an increased disorder strength, the electrical conductivity $\sigma$ drops as electronic states become gradually localized below the mobility edge. How fast $\sigma$ drops can also be related to the value of $x$ (whose exact value is not known) as shown in Eq.~(\ref{eq:4.2}); so, assuming that a large value $x$ corresponds to situations when the disorder strength is large, leads to see why the maximum conversion efficiency increases with disorder and $x$. This interpretation is in line with that in Ref.~\cite{yamamoto2017thermoelectricity}: \emph{increasing the disorder generates
more efficient thermoelectricity}. Finally, from a thermodynamic viewpoint, we may also relate this ascertainment to the fluctuation-compressibility theorem \cite{Kubo1965,Huang1986} mentioned in Ref.~\cite{khomchenko2022} where fluctuating Cooper pairs and nematic fluctuations were discussed: increasing disorder generates density fluctuation of the electronic extended states across the system, which in turn fosters the increase of the isentropic expansion factor $\gamma$ and hence of the maximum efficiency $\eta_{\rm max}$ defined in Eq.~(\ref{eq:5.1}).

\section{Conclusion}
We have connected the thermoelectric and thermoelastic properties of the three-dimensional Anderson model, discussing the role of the critical exponent. In contrast with the sharp enhancement of thermoelectric conversion close to the superconducting phase transition~\cite{khomchenko2022,ouerdane2015enhanced}, our results show a smooth, monotonously growing dependence of the thermoelectric figure of merit on temperature. Indeed, in single particle models a sharp energy-dependence of the transmission function is requested to obtain large thermoelectric efficiencies. On the other hand, in the Anderson model the dependence of the transmission function $\Sigma(E)$ on energy is non-analytical at the mobility edge $E_{\rm c}$ $-$ see  Eq.~(\ref{eq:4.3}) $-$, but less sharp than the more thermoelectrically efficient boxcar-function-shaped transmission functions~\cite{whitney}. However, as noted in \cite{yamamoto2017thermoelectricity}, though a large $ZT$ fosters a high energy conversion efficiency, this does not necessarily result in a high output power \cite{Apertet2012,Ouerdane2015}; so an optimal thermoelectric energy conversion in disordered electronic systems must account for the disorder strength when considering the power-efficiency trade-off.

Another essential issue is the influence of the mobility edge $E_{\rm c}$ on the thermoelectric properties of the Anderson transition. For simplicity, we considered $E_{\rm c}=0$, although one can choose a temperature-dependent mobility edge as in~\cite{premachandran1984temperature}. Such a choice may help to study the temperature-dependent interplay between the localized and extended states and how it may influence the electron gas figure of merit $z_{\rm e}T$. In~\cite{yamamoto2017thermoelectricity}, several values of $E_{\rm c}$ were considered and high values of the figure of merit $zT$ were obtained. A more realistic model of the mobility edge and how it may affect the energy conversion in the metal-insulator transition deserves more attention and is beyond of scope of the present work.  

On a general perspective, our results confirm the validity of the thermodynamic approach as a useful and physically intuitive way to estimate the ideal thermoelectric performance of the working fluid, neglecting the detrimental effect of phonons. Such an approach naturally suggests the consideration of electronic phase transitions to boost thermoelectric efficiency~\cite{khomchenko2022, ouerdane2015enhanced}.

\end{document}